\input ./style/arxiv-general.cfg
\documentclass[aoas,MSNbibl,nameyear,dvips]{arximspdf}
\makeatletter
   \@ifpackageloaded{graphicx}{}{\usepackage{graphicx}}
\makeatother

\doi{10.1214/15-AOAS863}
\volume{9}
\issue{4}
\pubyear{2015}
\firstpage{2023}
\lastpage{2050}
\docsubty{FLA}

\makeatletter

\newtheorem{satz}{Satz}[section]

\newtheorem{theo}[satz]{Theorem}
\newproclaim{ex}[satz]{Example}

\newproclaim{rem}[satz]{Remark}

\newproclaim{defi}[satz]{Definition}
\newcommand{\eqref}[1]{(\ref{#1})}
\newcommand{\R}{\mathbb R}
\newcommand{\E}{\mathbb E}
\newcommand{\N}{\mathbb N}
\renewcommand{\P}{\mathbb P}
\newcommand{\D}{ d}
\newcommand{\Zbf}{\mathbf Z}
\newcommand{\Xbf}{\mathbf X}
\newcommand{\Ybf}{\mathbf Y}
\newcommand{\tbf}{\mathbf t}
\newcommand{\T}{\mathrm{T}}
\newcommand{\SPEC}{\operatorname{SPEC}}
\newcommand{\CENS}{\operatorname{CENS}}
\newcommand{\HR}{\operatorname{HR}}
\makeatother

\begin{document}
\begin{frontmatter}

\title{Extremes on river networks\thanksref{T1}}
\runtitle{Extremes on river networks}
\thankstext{T1}{Supported by the Swiss National Science
Foundation and the EU-FP7 project Impact2C.}

\begin{aug}
\author[A]{\fnms{Peiman} \snm{Asadi}\thanksref{m1}\ead[label=e2]{peiman.asadi@unil.ch}},
\author[B]{\fnms{Anthony C.} \snm{Davison}\corref{}\thanksref{m2}\ead[label=e3]{anthony.davison@epfl.ch}}
\and\\
\author[B]{\fnms{Sebastian} \snm{Engelke}\thanksref{m1,m2}\ead[label=e1]{sebastian.engelke@epfl.ch}}
\runauthor{P.~Asadi, A.~C. Davison and S.~Engelke}
\affiliation{Universit\'e de Lausanne\thanksmark{m1} and Ecole
Polytechnique F\'ed\'erale de Lausanne\thanksmark{m2}}

\address[A]{P.~Asadi\\
Facult\'e des Hautes Etudes Commerciales\\
Universit\'e de Lausanne\\
Extranef, UNIL-Dorigny\\
1015 Lausanne\\
Switzerland\\
\printead{e2}}
\address[B]{A.~C. Davison\\
S.~Engelke\\
Ecole Polytechnique F\'ed\'erale de Lausanne\\
EPFL-FSB-MATHAA-STAT\\
Station 8\\
1015 Lausanne\\
Switzerland\\
\printead{e3}\\
\phantom{E-mail:\ }\printead*{e1}}
\end{aug}

%
\received{\smonth{2} \syear{2015}}
%
\revised{\smonth{7} \syear{2015}}

%
\begin{abstract}
Max-stable processes are the natural extension of the classical
extreme-value distributions to the functional setting, and they are
increasingly widely used to estimate probabilities of complex extreme
events. In this paper we broaden them from the usual situation in which
dependence varies according to functions of Euclidean distance to
situations in which extreme river discharges at two locations on a
river network may be dependent because the locations are flow-connected
or because of common meteorological events. In the former case
dependence depends on river distance, and in the second it depends on
the hydrological distance between the locations, either of which may be
very different from their Euclidean distance. Inference for the model
parameters is performed using a multivariate threshold likelihood,
which is shown by simulation to work well. The ideas are illustrated
with data from the upper Danube basin.
\end{abstract}

%
\begin{keyword}
\kwd{Extremal coefficient}
\kwd{hydrological distance}
\kwd{max-stable process}
\kwd{network dependence}
\kwd{threshold-based inference}
\kwd{upper Danube basin}
\end{keyword}
\end{frontmatter}

\section{Introduction}\label{sec1}

Modeling extreme events has recently become of great interest. The
financial crisis, heat waves, storms and heavy precipitation
underline the importance of assessing rare phenomena when few relevant
data are available.

There is a vast literature on modeling the univariate upper tail of the
distribution of environmental quantities such as precipitation or river
discharges at a fixed location $t$.
If $X_i(t)$ $( i=1,\ldots, n)$ are $n\in\N$ independent measurements of
a random spatial
process $X$ at location $t$, then the probability law of the maximum of
the~$n$ observations can be approximated by the generalized extreme
value distribution (GEVD)
%
\begin{equation}
\label{gevd} \P \biggl\{\max_{ i = 1,\ldots,n} \frac{X_i(t) -b_n} {a_n} \leq x
\biggr\} \approx G(x) = \exp \bigl\{ - (1 + \xi x )_+^{-1/\xi} \bigr\},\qquad
x\in \R,
\end{equation}
where $z_+ = \max(z,0)$ and $b_n \in\R$, $a_n > 0$ and $\xi\in\R$ are
the location, scale and shape parameters, respectively.
For $\xi= 0 $, $G(x)$ is read as the limit 
$\exp \{-\exp (-x ) \}$.
In fact, \eqref{gevd} represents the only possible nondegenerate limit
for maxima of independent and identically distributed sequences of
random variables [see, e.g., \citet{col2001}, Chapter~3].
This justifies the extrapolation to high quantiles using the parametric
tail approximation \eqref{gevd} for $u$ close to the upper endpoint of
the distribution of $X(t)$ by
%
\begin{equation}
\label{exceed_pareto} \P \bigl\{ X(t) > u \bigr\} \approx\frac{1}n \biggl(1 + \xi
\frac{u -
b_n}{a_n} \biggr)_+^{-1/\xi}.
\end{equation}

Often, however, univariate considerations are insufficient, because
near-simultaneous extreme events may cause the most severe damage. In
considering flooding of a river basin, for example, it is crucial to
understand the \emph{extremal dependence} between flows at different
gauging stations. Many authors have analyzed this using multivariate
copulas or multivariate extreme value distributions
[e.g., \citet{sal2010,ren2007}], but the explosion of the number of
parameters in high dimensions limits the applicability of such models,
and information on the geographical location of the stations cannot be
readily incorporated. Meteorological considerations suggest that
extremal dependence can be modeled as a function of the distance
between two locations. Indeed, for precipitation, temperature or wind
data, the use of \emph{Euclidean distance} has become standard in
spatial extremes [e.g., \citet{dav2012,hus2013,eng2014}]. An
important class of probability models for extreme spatial dependence on
the Euclidean space $\R^2$ is the class of \emph{max-stable
processes}, giving several flexible models whose dependence is
parameterized in terms of covariance functions [\citet{sch2002,opi2013}]
or of negative definite kernels [\citet{bro1977,kab2009,kab2011}]. Almost
all such models have hitherto presupposed that extremal dependence
depends only on the Euclidean distance between two locations, but this
may be too restrictive when more is known about the physical processes
underlying the data: locations on a river network may interact because
of the flow of water downstream between them.

\begin{figure}

\includegraphics{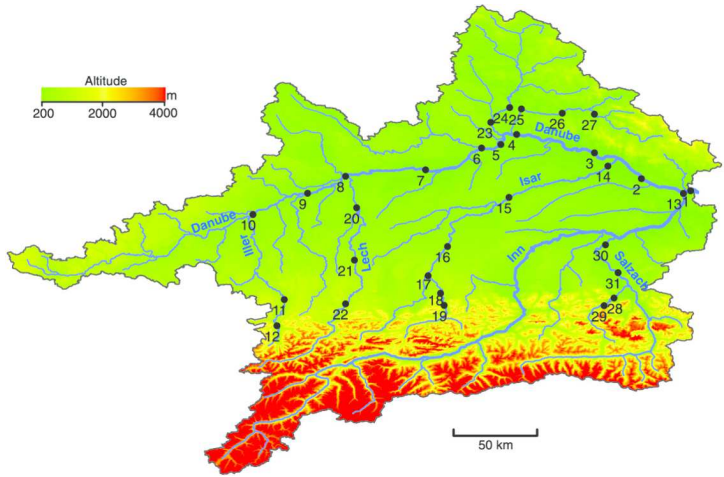}

\caption{Topographic map of the upper Danube basin, showing sites of
31 gauging stations (black circles) along the Danube and its
tributaries. Water flows toward gauging station 1.}
\label{Fig_Map}
\end{figure}

In this paper we focus on assessment of the risk of extreme discharges
on river networks in order to understand and prevent flooding. There is
longstanding interest in the application of extreme value statistics in
hydrology [e.g., \citet{kat2002,kee2009a,kee2009}]. In Europe, floods
are major natural hazards that can end human lives and cause huge
material damage. Figure~\ref{Fig_Map} shows the upper Danube basin,
which covers most of the German state of Bavaria and parts of Baden-W\"
urtemberg, Austria and Switzerland, and is regularly affected by
flooding. For this reason there is a well-developed system of gauging
stations that measure the daily average river discharge on this river
network; the locations of $31$ stations are shown on the map. For each
fixed location $t_j$ ($j=1,\ldots, 31)$ on the network, the
approximation \eqref{exceed_pareto} can be applied to daily measurements
$X_i(t_j)$ $(i=1,\ldots, n)$ of river discharge (m$^3$/s) in order to
model univariate tail probabilities.

Dependence modeling is more challenging. The extremal coefficient\break
$\theta(t_i,t_j)\in[1,2]$ measures the degree of dependence of large
values at two locations $t_i$ and $t_j$ on the river network; it ranges
from $\theta(t_i,t_j)=1$ for complete dependence to $\theta(t_i,t_j)=2$
for independence. The left panel of Figure~\ref{ECF_Euc} shows its
values for all pairs of stations in Figure~\ref{Fig_Map}, plotted
against their Euclidean distances. Unlike similar plots for extreme
precipitation, the non-Euclidean structure of the network means that
this graph shows only a weak relationship.

In this paper we aim to exploit both the geographical structure of the
river basin and the hydrological properties of the network in order to
provide a parsimonious model for extremal dependence. The resulting
dependence function has two parts:
\begin{itemize}
\item
since precipitation is the major source of extreme river discharges
and it is spatially dependent, one also expects higher
dependence of river discharges at stations which are close.
The left panel of Figure~\ref{ECF_Euc} suggests that the Euclidean distance
between stations has low explanatory power, so we shift each gauging station
to a new position in the center of its sub-catchment, which we call
its \emph{hydrological position}. The extremal coefficients plotted
against the \emph{hydrological distance} between the hydrological
positions exhibit a strong functional relationship, shown in the right
panel of Figure~\ref{ECF_Euc}, which is exploited in the dependence
model described in Section~\ref{sec_euc_dep};

\begin{figure}[t]

\includegraphics{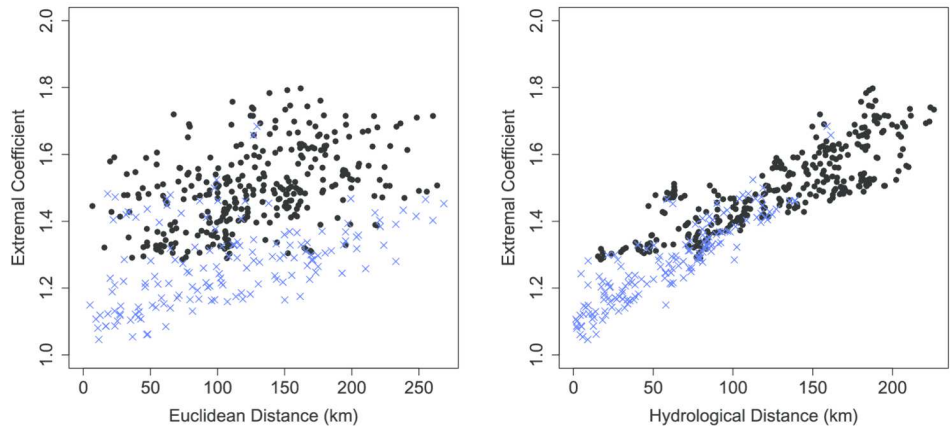}

\caption{Extremal coefficients (estimated using the madogram) of all
pairs of gauging stations plotted
against Euclidean distance (left) and hydrological distance (right);
those for
flow-connected pairs are blue crosses, and those for flow-unconnected
pairs are black circles.}
\label{ECF_Euc}
\end{figure}

\item
the crosses in Figure~\ref{ECF_Euc} represent the extremal
coefficients of pairs of flow-connected stations, which have
one station located upstream of the other.
Such pairs are generally more dependent
than flow-unconnected pairs, not only because the catchments are close
but also owing to the flow of water along the river. In Section~\ref
{sec_cov} we explain how
knowledge about the network structure and river sizes can be
included in the dependence model for flooding using ideas of \citet
{ver2010}, who defined covariance functions on river networks.
\end{itemize}

As one application of such a model, we would like to be able to compute
the multivariate counterpart of \eqref{exceed_pareto}, that is, the
probability of a rare event such as
\[
\P\bigl\{X(s_1) > u_1, \ldots,X(s_k) >
u_k\bigr\}
\]
for large $u_1,\ldots,u_k > 0$, where $s_1,\ldots, s_k \in T$ can be any
stations on the river network, even without measurements there. More
complicated quantities, such as the sum of discharges at several
stations, may also be of interest.

\section{Preliminaries}
\subsection{Extreme value theory}
\label{sec_EVT}
The only nontrivial limiting distribution for the normalized maxima of
an independent and identically distributed sequence of scalar random
variables is the max-stable GEVD, expression \eqref{gevd}. In the
multivariate case, we can transform each margin such that the max-limit
has a standard Fr\'echet cumulative distribution function $\exp(-1/x)$
$(x > 0)$. In this way, without loss of generality, we can concentrate
on the multivariate dependence between the components
[\citet{res2008}, Proposition~5.8].

Let $\Xbf_i = (X_{1,i}, \ldots,X_{m,i})$ $(i = 1,\ldots,n)$ be
independent copies
of an $m$-variate random vector $\Xbf$ and assume that for each
$j=1,\ldots,m$
the maximum $\max_i X_{j,i}$ converges
to a GEVD $G_j$, as in \eqref{gevd}, with norming constants
$b_{j,n}\in
\R$, $a_{j,n}>0$
and shape parameter $\xi_j$. Define the transformations
%
\begin{equation}
U_{j}(x) = - 1 / \log G_j (x ) = ( 1 +
\xi_j x )_+^{1/{\xi_j}}, \label{U_transform}
\end{equation}
and note that
\[
\lim_{n\to\infty} \P \biggl\{\max_{i = 1,\ldots, n}
U_{j} \biggl(\frac
{X_{j,i} - b_{j,n}}{a_{j,n}} \biggr) \leq x \biggr\} = \exp(-1/x),\qquad j
= 1,\ldots, m.
\]

We say that $\Xbf$ is in the multivariate maximum domain of attraction
(MDA) of a random vector
$\Zbf= (Z_1,\ldots, Z_m)$, if for any $\mathbf z= (z_1,\ldots, z_m)$,
%
\begin{eqnarray}
\label{X_conv} &&\lim_{n\to\infty} \P \biggl\{\max_{i = 1,\ldots, n}
U_{1} \biggl(\frac
{X_{1,i} - b_{1,n}}{a_{1,n}} \biggr) \leq z_1, \ldots,
\max_{i =
1,\ldots, n} U_{m} \biggl(\frac{X_{m,i} - b_{m,n}}{a_{m,n}} \biggr)
\leq z_m \biggr\}
\nonumber
\\[-8pt]
\\[-8pt]
\nonumber
&&\qquad = \P ( \Zbf\leq\mathbf z );
\end{eqnarray}
call this joint distribution $F_\Zbf(\mathbf z)$. In this case, $\Zbf
$ is
max-stable with standard Fr\'echet marginal distributions; see before
\eqref{Z_conv}. Moreover, by \citet{res2008}, Proposition~5.8,
we may write
%
\begin{equation}
\label{exp_measure} F_\Zbf(\mathbf z) = \exp\bigl\{ - V(\mathbf z)\bigr\},\qquad
\mathbf z\in\R^m,
\end{equation}
where the exponent measure $V$ is a measure defined on the cone $E =
[0,\infty)^m\setminus\{\mathbf{0}\}$ and $V(\mathbf z)$ is shorthand
for $V([\mathbf{0},\mathbf z]^C)$. The object $V$ incorporates the extremal
dependence structure of $\Zbf$, where $V(\mathbf z) = 1 / \min
(z_1,\ldots,\break z_m)$ and $V(\mathbf z) = 1 / z_1 + \cdots+ 1/z_m$ represent complete
dependence and independence, respectively.
The measure $V$ is homogeneous of order $-1$, that is, $V(\lambda
\mathbf z)
= \lambda^{-1} V(\mathbf z)$, for $\lambda>0$,
and it satisfies $V(z,\infty,\ldots,\infty) = 1 / z$ for $z > 0$ and any
permutation of its arguments.
There are many parametric models for the exponent measure $V$ and thus
for multivariate extreme value distributions or copulas.
The explosion of parameters in most such models makes fitting them
feasible only in low dimensions.

By Proposition~5.17 of \citet{res2008} the convergence in \eqref{X_conv}
is equivalent to
%
\begin{equation}
\label{X_conv_meas} \lim_{n\to\infty} n \P \biggl[ \biggl\{U_{1}
\biggl(\frac{X_{1} -
b_{1,n}}{a_{1,n}} \biggr), \ldots, U_{m} \biggl(
\frac{X_{m} -
b_{m,n}}{a_{m,n}} \biggr) \biggr\} \in A \biggr] = V(A)
\end{equation}
for any Borel subset $A \subset E$ which is bounded away from $\mathbf
0$ and satisfies $V(\partial A) = 0$, where $\partial A$ is the
boundary of $A$. This important observation allows us to approximate
the probability that $\Xbf$ falls into a rare region. For instance, if
$A = (u_1,\infty)\times\cdots\times(u_m,\infty)$ $(u_1,\ldots,u_m
\in
\R)$, then for large $n$ \eqref{X_conv_meas} implies that
%
\begin{equation}
\label{mult_exceed} \P ( X_1 > u_1, \ldots, X_m
> u_m ) \approx\frac{1}n V \Biggl\{ \prod
_{j=1}^m \biggl( U_{j} \biggl(
\frac
{u_j -
b_{j,n}}{a_{j,n}} \biggr), \infty \biggr) \Biggr\},
\end{equation}
where $\prod$ denotes the Cartesian product.
More complicated events such as $A = \{ \mathbf x\in\R^m: \sum_{i=1}^m x_i
> u\}$ for some $u\in\R$ can also be considered. Equation \eqref
{X_conv_meas} implies that as $n\to\infty$ the empirical point process
\[
\biggl\{ \biggl(U_{1} \biggl(\frac{X_{1,i} - b_{1,n}}{a_{1,n}} \biggr), \ldots,
U_{m} \biggl(\frac{X_{m,i} - b_{m,n}}{a_{m,n}} \biggr) \biggr): i=1,\ldots, n \biggr\}
\]
converges vaguely to a Poisson point process on $E$ with intensity
measure $V$ [\citet{res2008}, Proposition~3.21]. In Section~\ref
{inference} this result will be used to derive the asymptotic
distribution of exceedances and to fit parametric models for~$V$.

In the bivariate case $m=2$, a common summary statistic for the
dependence among components of $F_\Zbf$ is the extremal coefficient
$\theta\in[1,2]$ [see, e.g., \citet{sch2003}], which is defined
through the expression
%
\begin{equation}
\label{EC} \P(Z_1\leq u, Z_2 \leq u) =
\P(Z_1\leq u)^\theta, \qquad u>0,
\end{equation}
or, equivalently, $\theta= V(1,1)$. Consequently, the cases $\theta=
1$ and $\theta=2$ correspond to complete dependence and independence.
Model-free estimation of the extremal coefficient is possible through
the madogram [\citet{coo2006}], and these estimates of $\theta$ can be
used for model-checking.

\subsection{Max-stable processes}
\label{processes}

Max-stable processes can be defined on any index set $T$, though this
is usually taken to be a subset of an Euclidean space $\R^d$. A random
process $\{ Z(t):t\in T\}$ is called \emph{max-stable} if there exists
a sequence $(X_i)_{i\in\N}$ of independent copies of a process $\{
X(t):t\in T\}$ and functions $a_n(t) > 0$, $b_n(t)\in\R$, such that
the convergence
%
\begin{equation}
\label{Z_conv} Z(t) = \lim_{n\to\infty} \Bigl\{\max
_{i=1,\ldots, n} X_i(t) - b_n(t) \Bigr
\}/a_n(t),\qquad t\in T,
\end{equation}
holds in the sense of finite dimensional distributions. In this case,
the process $X$ is said to lie in the max-domain of attraction of $Z$.

The class of max-stable processes is generally too large for
statistical modeling, so one typically considers parametric subclasses
of models. Examples include mixed moving maxima processes [\citet
{wan2010}], Schlather processes [\citet{sch2002}] and Brown--Resnick
processes [\citet{bro1977,kab2009}]. In this paper we rely on the
construction principle for a large class of max-stable processes given
in \citet{kab2011}; see also \citet{kab2009}. A \emph{negative definite
kernel} $\Gamma$ on an arbitrary nonempty set $T$ is a mapping $\Gamma:
T\times T \to[0,\infty)$ such that for any $n\in\N$ and $a_1,\ldots,
a_n \in\R$ with $\sum_{i=1}^n a_i = 0$, we have
\[
\sum_{i=1}^n \sum
_{j=1}^n a_i a_j
\Gamma(t_i,t_j) \leq0, \qquad t_1,
\ldots,t_n\in T.
\]
The following result states that there corresponds a max-stable process
to any negative definite kernel on $T$.

\begin{theo}[{[\citet{kab2011}, Theorem~1]}]
\label{thm1}
Suppose that $W_i$ ($i\in\N$) are independent copies
of the zero-mean Gaussian process $\{W(t): t\in T\}$
whose incremental variance
$\E\{ W(s) - W(t) \}^2$ equals $\Gamma(s,t)$ for all
$s,t\in T$. Let $\sigma^2(t) = \E\{W(t)^2\}$ denote the variance
function of $W$ and let $\{ U_i: i\in\N\}$ denote a Poisson process on
$(0,\infty)$ with intensity $u^{-2} \,du$.
Then the process
%
\begin{equation}
\label{BR_def} \eta_\Gamma(t) = \max_{i\in\N}
U_i \exp\bigl\{W_i(t) - \sigma^2(t)/2\bigr
\},\qquad  t\in T,
\end{equation}
is max-stable, has standard Fr\'echet margins, and its distribution
depends only on~$\Gamma$.
\end{theo}

If $T=\R^d$ and $W$ is an intrinsically stationary Gaussian process,
then $\eta_\Gamma$ is called a Brown--Resnick process [\citet{bro1977,
kab2009}]. This is a popular model for complex extreme events. The
generation of random samples from Brown--Resnick type processes
is challenging [cf. \citet{eng2011, oes2012}], but recent advances
provide exact and efficient algorithms [\citet{die2015, dom2015}].

\begin{rem}
\label{rem_BR}
\textup{(a)}
For any negative definite kernel $\Gamma$ there are many different
Gaussian processes with incremental variance $\Gamma$ [\citet
{kab2011}, Remark~1]. In particular,
for $u\in T$, we can choose a unique Gaussian process $W^{(u)}$
with incremental variance $\Gamma$ and $W^{(u)}(u) = 0$ almost surely. The
covariance function of this process is
%
\begin{equation}
\label{gamma_cov} \E\bigl\{ W^{(u)}(t) W^{(u)}(s)\bigr\} = \bigl\{
\Gamma(s,u) + \Gamma(t,u) - \Gamma(s,t)\bigr\}/2.
\end{equation}
Thus, there is a one-to-one correspondence between negative definite
kernels $\Gamma$ and the class of max-stable processes $\eta_\Gamma$.\vspace*{-6pt}
\begin{longlist}[(b)]
\item[(b)]
If $\{X(t): t\in T\}$ is a zero-mean
Gaussian process with covariance function $C:T\times T\to\R$,
then $\Gamma(s,t) = C(s,s) + C(t,t) - 2C(s,t)$ is a negative definite
kernel on $T$.
\end{longlist}
\end{rem}

The bivariate distribution function of $(\eta_{\Gamma}(s), \eta
_{\Gamma
}(t))$ $(s,t\in T)$
is
%
\begin{eqnarray}\label{biv_cdf}
&&
\P \bigl\{\eta_{\Gamma}(s)\leq x, \eta_{\Gamma}(t) \leq y
\bigr\}
\nonumber\\
&&\qquad=\exp \biggl\{-\frac{1}x\Phi \biggl[\frac{\sqrt
{\Gamma
(s,t)}}{2} +
\frac{\log(y/x)}{\sqrt{\Gamma(s,t)}} \biggr] -\frac{1}y\Phi \biggl[\frac{\sqrt{\Gamma(s,t)}}{2} +
\frac{\log
(x/y)}{\sqrt{\Gamma(s,t)}} \biggr] \biggr\},\\
\eqntext{ x, y > 0,}
\end{eqnarray}
where $\Phi$ is the standard normal distribution function. Analogously
to the extremal coefficient in \eqref{EC}, one considers the \emph
{extremal coefficient function} $\theta(s,t)$ $(s,t \in T)$, defined as
the extremal coefficient of the bivariate vector $(\eta_\Gamma(s),
\eta
_\Gamma(t))$, as a measure of the functional extremal dependence of the
max-stable process $\eta_\Gamma$.
By~\eqref{biv_cdf}, we conclude that
%
\begin{equation}
\label{ECF} \theta(s,t) =
 2 \Phi \biggl\{\frac{\sqrt{\Gamma(s,t)}}{2} \biggr\},
\end{equation}
so the negative definite kernel $\Gamma$ parameterizes the extremal
dependence between
observations at positions $s$ and $t$; small and large values of
$\Gamma(s,t)$
correspond to strong and weak dependence, respectively. By Remark~\ref
{rem_BR}(a), any kernel $\Gamma$ yields a max-stable process $\eta
_\Gamma$, so in Section~\ref{sec_model} we can and will focus on
finding a parametric model for $\Gamma$ suitable for our application.

The higher dimensional distributions of $\eta_\Gamma$ are more complicated.
For instance, for $\tbf= (t_1,\ldots, t_m)\in T^m$,
the random vector $(\eta_\Gamma(t_1), \ldots, \eta_\Gamma(t_m))$ is
max-stable and its exponent measure $V_{\Gamma,\tbf}$ defined in
\eqref
{exp_measure}
is characterized by [\citet{kab2011}]
%
\begin{equation}
\label{HR} V_{\Gamma,\tbf}(x_1,\ldots, x_m) = \E
\biggl[ \max_{i=1,\ldots,m} \biggl\{ \frac{W(t_i) - \sigma^2(t_i)/2}{x_i} \biggr\} \biggr].
\end{equation}
This multivariate max-stable distribution is called the H\"usler--Reiss
distribution [\citet{hue1989}].
Computation of the expected value in \eqref{HR} involves
high-dimensional integrals and thus is awkward in general.

\section{Model}
\label{sec_model}

\subsection{River network}
\label{sec_network}

In the previous section we showed how to define max-stable processes on
an arbitrary index set $T$. From here on, $T$ will represent a river
network and we will construct a kernel $\Gamma$ flexible enough to
explain the extremal dependence observed in data.

\begin{figure}

\includegraphics{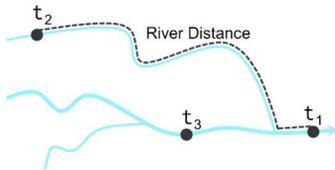}

\caption{River network with three locations $t_1,t_2,t_3\in T$; $t_1$
is flow-connected
with both $t_2$ and $t_3$, but $t_2$ and $t_3$ are flow-unconnected.}
\label{Fig_RiverDist}
\end{figure}

Let us first fix some notation for river networks [\citet{ver2010}]. We
embed our network $T$ in the Euclidean space $\R^2$ representing the
geographical river basin. To this end, let $T\subset\R^2$ denote the
collection of piecewise differentiable curves, called river segments,
that are connected at the junctions of the river and whose union
constitutes the river network. There is a finite number $M \in\N$ of
such segments and we index them by $i \in\mathcal S = \{1, \ldots, M\}
$. The network is dendritic, in the sense that there is one most
downstream segment, which splits up into other segments when going
upstream; see Figure~\ref{Fig_RiverDist}. For a location $t_i \in T$ on
the $i$th segment, we let $D_i\subseteq\mathcal S$ denote the index
set of river segments downstream of $t_i$, including the $i$th segment.
Moreover, for another location $t_j\in T$ on the $j$th segment we say
that $t_i$ and $t_j$ are \emph{flow-connected}, written $t_i
\leftrightarrow t_j$, if and only if $D_i\subseteq D_j$ or
$D_j\subseteq D_i$. If $t_i$ and $t_j$ are not flow-connected, we say
that they are \emph{flow-unconnected} and write $t_i \nleftrightarrow
t_j$. If $t_j$ is \emph{upstream} of $t_i$, that is, $D_i\subset D_j$,
then we denote the set of segments between $t_j$ and $t_i$, inclusive
of the $j$th but exclusive of the $i$th segment, by $B_{i,j} = D_j
\setminus D_i$. If $t_j$ is downstream of $t_i$, then $B_{i,j} = D_i
\setminus D_j$. In the case that $t_i$ and $t_j$ are on the same
segment, that is, $D_i = D_j$, we put $B_{i,j} = \varnothing$.

\begin{figure}

\includegraphics{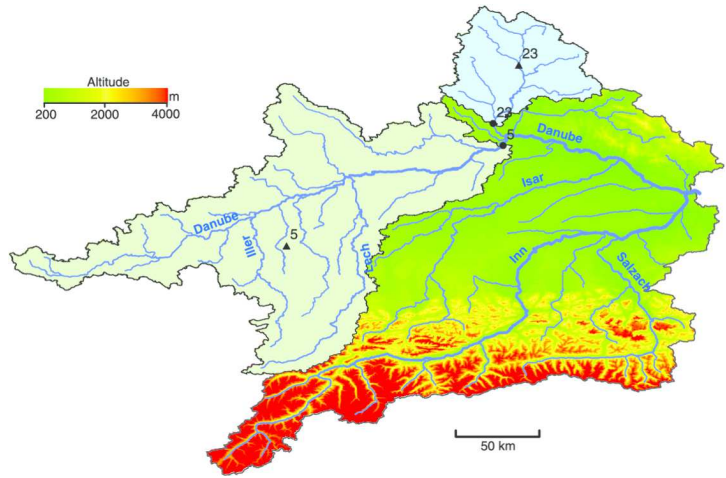}

\caption{Gauging stations 5 and 23 (black circles), their
sub-catchments in light green and blue, respectively,
and their hydrological locations (black triangles) as defined
in \protect\eqref{centroid}.}
\label{Fig_Subcatch}
\end{figure}

We define the \emph{river distance} $d(t_1,t_2)$ between two arbitrary
points $t_1,t_2$ on the network $T$
as the shortest distance along $T$, that is, we sum the arc-lengths of
the segment curves lying between $t_1$ and $t_2$; see Figure~\ref
{Fig_RiverDist}. The embedding of the river network $T$ in
the Euclidean space $\R^2$ has the advantage that
we can exploit the geographical structure of the river basin.
To this end, associate to each location $t = (x,y) \in T \subset\R^2$
the set $S_t \subset\R^2$ of all points on the
geographical map such that water from this point will eventually flow
through point $t$ on the river. The set $S_t$ is called the sub-catchment
of location $t$; see Figure~\ref{Fig_Subcatch}.

As explained in Section~\ref{processes},
we need to construct a negative definite kernel $\Gamma$
on the space $T\times T$ that captures the dependence
structure of extreme values on the river network $T$.
Figure~\ref{ECF_Euc} suggests that this should be based on two
components: one, $\Gamma_{\mathrm{Riv}}$, for the flow-connected dependence
along the river, taking into account the hydrological properties of the
river network; and another, $\Gamma_{\mathrm{Euc}}$, for the
dependence resulting from the geographical structure
of the river basin and spatially distributed meteorological
variables.

\subsection{Dependence measure $\Gamma_{\mathrm{Riv}}$}
\label{sec_cov}

There are many models for Gaussian random fields
where the covariance between two locations depends only on
the Euclidean distance between two points. Such covariances are not
valid with metrics such as the
river distance $d$ on our network because they may not be positive
definite. Recent work [\citet{ver2006,cre2006,ver2010}] has developed
covariances that are positive definite as functions of river distance.
A related approach, the top-kriging
of \citet{sko2006}, uses variograms integrated over catchments, but
does not provide closed-form formulae, so we focus on river
distance methods.

Following the ``upstream construction'' in \citet{ver2006}, we can
define a covariance function based on river distance for $t_i,t_j\in T$ by
%
\begin{eqnarray}
\label{cov_stream} C_{\mathrm{Riv}}(t_i,t_j) = \cases{
\displaystyle\prod_{k\in B_{i,j}} \sqrt{\pi_k} C_1
\bigl\{d(t_i,t_j)\bigr\}, &\quad $t_i
\leftrightarrow t_j$, \vspace*{2pt}
\cr
0, &\quad $ t_i
\nleftrightarrow t_j$, }
\end{eqnarray}
where the covariance function $C_1$ arises from a moving average
construction on~$\R$.
If $B_{i,j} = \varnothing$ in \eqref{cov_stream}, then $\prod_{k\in
B_{i,j}} \sqrt{\pi_k}$
is set to $1$. The corresponding
weights $\pi_k$ ($k\in B_{i,j}$) are chosen such that the
variance is constant, that is, $C_{\mathrm{Riv}}(t_i,t_i) = C_{\mathrm{Riv}}(t_j,t_j) = C_1(0)$ for all
$t_i,t_j\in T$. For a fuller treatment, see \citet{ver2006} and \citet
{ver2010}, who
also provide different parametric classes for the covariance
function $C_1$, including the \emph{linear with sill model}
\[
C_1(h) = ( 1 - h/\tau )_+, \qquad \tau> 0,
\]
which we use below.
Intuitively, the covariance function \eqref{cov_stream}
can be understood as follows: an event at a downstream location,
for example, $t_1$ in Figure~\ref{Fig_RiverDist}, can be caused
by an event on one of the two branches of an upstream bifurcation. The
weights $\pi_k$ quantify the proportions of events coming
from the branches. If several bifurcations lie between
two flow-connected locations, then the weights along the connection
must be multiplied.
The choice of the weights in the covariance function $C_{\mathrm{Riv}}$ in
\eqref{cov_stream} is crucial and depends on the application. As we
consider extreme discharges on river networks, the weights at a
bifurcation should reflect the proportion of large discharge values at
the downstream river that are caused by a large discharge of one of the
upstream rivers. In Figure~\ref{Fig_RiverDist}, for example, a natural
choice for the weights $\pi_2$, $\pi_3$ on the river segments of $t_2$,
$t_3$ is to take the proportion of mean water volumes, that is, $\pi
_{i} = E_{t_i} / ( E_{t_2} + E_{t_3})$, where $E_{t_i}$ is the average
discharge at location $t_i$ $(i=2,3)$. This, however, requires
measurements at all bifurcations. Since we would like to use our model
for extrapolation to parts of the network without measurements, we must
approximate $E_{t_1}$ and $E_{t_2}$. A digital elevation model can be
used to extract the geographical coordinates of the sub-catchment $S_t$
corresponding to each location $t\in T$ on the river network, including
the altitude $h(x,y)$ at all $(x,y) \in S_t$. Exploratory analysis
shows that altitude is an excellent covariate for average
precipitation, so we define $E_t^*$ as the integrated altitude over
$S_t$, that is,
\[
E_t^* = \int_{S_t} h(x,y) \,\D x \,\D y,
\]
which is thus approximately proportional to the average runoff
accumulated in the sub-catchment $S_t$. We then define the weights in
the above example to be
%
\begin{equation}
\label{weights_def} \pi_{i} = E_{t_i}^* / \bigl(
E_{t_2}^* + E_{t_3}^*\bigr), \qquad i=2,3.
\end{equation}
By the second part of Remark~\ref{rem_BR} and the construction
of the positive definite covariance function in \eqref{cov_stream},
we obtain a negative definite
kernel $\Gamma_{\mathrm{Riv}}$ on the river network $T$ by setting
\begin{eqnarray*}
\Gamma_{\mathrm{Riv}}(t_i,t_j) = \cases{ 1 -\displaystyle \prod
_{k\in B_{i,j}} {\sqrt{\pi_k}} \bigl( 1 -
d(t_i,t_j)/\tau \bigr)_+, &\quad $t_i
\leftrightarrow t_j$, \vspace*{2pt}
\cr
1, &\quad $ t_i
\nleftrightarrow t_j$. }
\end{eqnarray*}

\subsection{Dependence measure $\Gamma_{\mathrm{Euc}}$}
\label{sec_euc_dep}

Two flow-unconnected locations on the river network can have dependent
extreme discharges, since precipitation is spatially dependent. As
shown in Figure~\ref{ECF_Euc}, the usual Euclidean distance between two
points cannot fully explain this dependence, because the total amount
of water at location $t\in T$ on the river network comes not only from
precipitation there, but also from the accumulated runoff from its
sub-catchment $S_t$. Thus, instead of the Euclidean distance between
two points $s,t \in T$, we should consider a \emph{hydrological
distance} that appropriately describes the distance between runoff in
sub-catchments $S_s$ and $S_t$ due to precipitation. For this purpose
we first shift each location $t\in T$ to a \emph{hydrological location}
by a function $H:T \to\R^2$. In our case, the center of mass of mean
annual precipitation on the sub-catchment $S_t$ gives a good choice
[\citet{mer2005}]. As noted in Section~\ref{sec_cov}, precipitation data
on a dense grid is often difficult to obtain, so we use the altitude
$h(x,y)$ at location $(x,y)\in S_t$ instead.

The hydrological location $H(t)$, or ``altitude weighted
centroid,'' of a point on the river network is
%
\begin{equation}
\label{centroid} H(t) = \biggl(\frac{1}{E_t^*} \int_{S_t} x
h(x,y) \,\D x \,\D y, \frac{1}{E_t^*} \int_{S_t} y h(x,y) \,\D x
\,\D y \biggr)^\T,\qquad t\in T,
\end{equation}
and the hydrological distance between $s,t \in T$ is $\| H(s) - H(t) \|
$, where $\| \cdot\|$ denotes Euclidean distance. Figure~\ref
{Fig_Subcatch} shows two stations on the river network that are close
in terms of Euclidean distance but whose hydrological locations are far
apart. The right-hand panel of Figure~\ref{ECF_Euc} reveals strong
functional dependence of the extremal coefficients on hydrological distance.

A variogram that is valid on the Euclidean space $\R^2$ can be applied
to the hydrological positions $H(t)$ ($t\in T$). The fractal variogram
family $\Gamma_\alpha(x,y) = \| x - y \|^\alpha$ $(x,y\in\R^2)$, where
$\alpha\in(0,2]$ is called the shape parameter, is commonly used, but
it is isotropic: the dependence decreases at the same rate in each
direction. Extremal meteorological data often exhibit anisotropies that
can be captured by including a rotation and dilation matrix [\citet
{bla2011,eng2014}]
%
\begin{equation}
\label{aniso_mat} R\equiv R(\beta, c) =\pmatrix{\cos\beta& -\sin\beta
\vspace*{2pt}
\cr
c \sin\beta& c\cos\beta},\qquad \beta\in[\pi/4,3\pi/4], c > 0,
\end{equation}
where the restriction of $\beta$ to one quadrant ensures the
identifiability of the parameters $(\beta, c)$. Applying the kernel
$\Gamma_\alpha$ and transformation $R$ to the positions $H(t)$, we
obtain a negative definite kernel on the river network $T$, that is,
\[
\Gamma_{\mathrm{Euc}}(t_i,t_j) = \bigl\| R\cdot
H(t_i) - R\cdot H(t_j) \bigr\|^\alpha ,\qquad
t_i,t_j\in T,
\]
where $R\cdot v$ denotes matrix multiplication of $R$ and the vector
$v\in\R^2$.

\subsection{Max-stable process on $T$}

In Sections \ref{sec_cov} and \ref{sec_euc_dep} we defined two negative
definite kernels on the river network $T$: $\Gamma_{\mathrm{Riv}}$ models
the extremal dependence of flow-connected stations due to the specific
hydrological properties of the river network, and $\Gamma_{\mathrm{Euc}}$
describes additional dependence between all stations due to the
geographical structure of the river basin and spatially distributed
precipitation. We combine these to obtain our final dependence model:
given weights
$\lambda_{\mathrm{Riv}}, \lambda_{\mathrm{Euc}} \geq0$, we put
%
\begin{eqnarray}
\label{dep_kernel} \Gamma(t_i,t_j) &=& \lambda_{\mathrm{Riv}}
\Gamma_{\mathrm{Riv}}(t_i,t_j) + \lambda _{\mathrm{Euc}}
\Gamma_{\mathrm{Euc}}(t_i,t_j)
\nonumber
\\[-8pt]
\\[-8pt]
\nonumber
&=& \cases{ \lambda_{\mathrm{Riv}} \displaystyle\biggl\{1 - \prod
_{k\in B_{i,j}} \sqrt{\pi_k} \bigl( 1 +
d(t_i,t_j)/\tau \bigr)_+ \biggr\} \vspace*{2pt}\cr
\quad{}+
\lambda_{\mathrm{Euc}} \bigl\| R\cdot H(t_i) - R\cdot H(t_j)
\bigr\|^\alpha, & \quad $t_i \leftrightarrow t_j$,
\vspace*{2pt}
\cr
\lambda_{\mathrm{Riv}} + \lambda_{\mathrm{Euc}} \bigl\| R\cdot
H(t_i) - R\cdot H(t_j) \bigr\|^\alpha,
&\quad $t_i \nleftrightarrow t_j$,}
\end{eqnarray}
for any $t_i,t_j\in T$.
By Remark~\ref{rem_BR} we can define
a Gaussian random field $W$ on $T$ with variogram $\Gamma$, and
by Theorem~\ref{thm1} we obtain a max-stable process $\eta_\Gamma$
on~$T$, defined in \eqref{BR_def}, with dependence
function $\Gamma$. The process $\eta_\Gamma$ is nonstationary: indeed,
since it is not defined on a Euclidean space, even the notion of
stationarity is unclear.

The process $\eta_\Gamma$ has standard Fr\'echet margins. However,
even after normalization of the data with scale and location parameters
at each location $t\in T$ as in~\eqref{gevd}, the univariate
tail distributions will have different shapes. We must therefore transform
the standard Fr\'echet margins in \eqref{BR_def} to GEVD. We set
%
\begin{equation}
\label{eta_tilde} \tilde{\eta}_\Gamma(t) = \frac{\eta_\Gamma(t)^{\xi(t)} - 1}
{\xi(t)},\qquad t\in T,
\end{equation}
where $\xi(t)\in\R$ is the shape parameter at point $t\in T$. It is
then easily verified that the margins of $\tilde\eta_\Gamma$ follow a
GEVD, that is,
\[
\P \bigl\{\tilde{\eta}_\Gamma(t) \leq x \bigr\} = \exp \bigl[ - \bigl
\{1 + \xi(t) x \bigr\}_+^{-1/\xi(t)} \bigr],\qquad x\in \R.
\]

\section{Inference}
\label{inference}

\subsection{General}
\label{sec_general}

Inference for the extremes of univariate data is well developed [\citet
{col2001, deh2006a, emb1997}], so we merely sketch it in Section~\ref
{sec_stat_uni}. Statistical inference for multivariate or spatial
models is more difficult, as their distributions are rarely known in
closed form or involve high-dimensional integration. Composite
likelihood methods based on bivariate densities have therefore been
widely applied [\citet{pad2010,dav2012,hus2013}]. Recent research has
focused on methods that exploit full likelihoods of multivariate
extreme observations through peaks-over-threshold approaches [\citet
{wad2013, eng2014, thi2014, bie2014}] and on $M$-estimators for spatial
extremes [\citet{ein2014}]. However, different definitions of an extreme
event yield different inferences. One might call a multivariate
observation extreme if at least one component is large, leading to
multivariate generalized Pareto distributions [\citet{roo2006}], whereas
choosing data where a single fixed component exceeds a high threshold
gives a conditional extreme value model [\citet{hef2004}], and spectral
estimation is based on observations where a suitable norm of the
components is large [cf. \citet{col1991}]. For finite samples each
choice has advantages and disadvantages [\citet{Huser.Davison.Genton:2014}].

We consider two estimation procedures tailor-made for a max-stable
process $\eta_\Gamma$ whose finite-dimensional margins follow the H\"
usler--Reiss distribution \eqref{HR}. \citet{eng2014} compute the
spectral density of the exponent measure \eqref{HR} and introduce an
estimator for the parameters of a Brown--Resnick process [\citet
{kab2009}]. \citet{wad2013} use events for which at least one component
exceeds a high threshold, and censor any components that stay below it.

In Section~\ref{sec_stat_mult} we review these two methods, show how
they can be adapted to our framework, and derive a new representation
of the conditional densities, simpler than that in \citet{wad2013}.
\citet{Asadi.Davison.Engelke:2016:SM} describe a small simulation study
that aids in the choice of estimator for our application.

\subsection{Univariate margins}
\label{sec_stat_uni}

We must estimate the univariate extreme value parameters, that is, the
norming constants $a_{j,n}$, $b_{j,n}$, and the shape parameter $\xi_j$
($j = 1,\ldots, m$) in \eqref{gevd}. This allows the calculation of
univariate return levels at each location and is needed for the
transformations $U_{j,n}$ in \eqref{U_transform} that appear in the
multivariate exceedance probabilities \eqref{mult_exceed}. We use the
Poisson point process approach [\citet{col2001}, Section~7.3] to fit
these models for the univariate exceedances.

Recall that $\Xbf_i = (X_{1,i}, \ldots,X_{m,i})$ $(i = 1,\ldots, n)$ are
independent copies of an $m$-variate random vector $\Xbf$ as in
Section~\ref{sec_EVT}. For each location $j=1,\ldots,m$, let $q_{j,p}$
be the empirical $p$-quantile, with $p\approx1$, of the data $X_{j,1},
\ldots,\break X_{j,n}$, and write $\mathcal I_j = \{i\in\{1,\ldots, n\}:
X_{j,i} > q_{j,p}\}$. Then the Poisson point process likelihood for the
exceedances at station $t_j$, assumed independent, can be written as
[\citet{col2001}, (7.9)]
%
\begin{eqnarray}
\label{PPP_density} L(\xi_j,a_{j,n},b_{j,n}) &\propto&
\exp \biggl\{ - n_j \biggl[1 + \xi_j \biggl(
\frac{q_{j,p} - b_{j,n}}{a_{j,n}} \biggr) \biggr]^{-1/\xi_j} \biggr\}
\nonumber
\\[-8pt]
\\[-8pt]
\nonumber
&&{}\times \prod
_{i \in\mathcal I_j} a_{j,n}^{-1} \biggl[ 1 +
\xi_j \biggl(\frac
{X_{j,i} - b_{j,n}}{a_{j,n}} \biggr) \biggr]^{-1/\xi_j -1},
\end{eqnarray}
where $n_j$ is the number of years of observations at location $t_j$.
Owing to the inclusion of $n_j$, the parameters $a_{j,n}, b_{j,n}$ and
$\xi_j$ equal those in the GEVD \eqref{gevd} for yearly maxima. A joint
model for the parameters at different locations, such as a linear model
with environmental covariates, can be fitted by maximizing a so-called
independence likelihood [\citet{Chandler.Bate:2007}] based on the product
of \eqref{PPP_density} over all stations.

\subsection{Estimation of \texorpdfstring{$\eta_\Gamma$}{$eta_Gamma$}}
\label{sec_stat_mult}

In order to fit the max-stable process $\eta_\Gamma$ introduced in
Section~\ref{sec_model} with dependence kernel \eqref{dep_kernel}, we
must estimate the six parameters
%
\begin{eqnarray}
\label{six_par} \lambda_{\mathrm{Riv}}&\geq&0,\qquad
 \lambda_{\mathrm{Euc}} \geq0,\qquad \tau>
0,
\nonumber
\\[-8pt]
\\[-8pt]
\nonumber
 \alpha&\in&(0,2], \qquad \beta\in[\pi/4,3\pi/4], \qquad c > 0,
\end{eqnarray}
that characterize the river and Euclidean dependence functions $\Gamma
_{\mathrm{Riv}}$ and $\Gamma_{\mathrm{Euc}}$ and their weights. Below we write
$ \vartheta= (\lambda_{\mathrm{Riv}}, \lambda_{\mathrm{Euc}}, \tau, \alpha,
\beta
, c)$, and denote the corresponding parameter space by $\Theta$.
When stressing that $\Gamma$ depends on the parameter~$\vartheta$, we
write $\Gamma= \Gamma_\vartheta$.

We do not observe data from the asymptotic limit model $\eta_\Gamma$
itself, so let us specify the assumptions for our observations. As in
Section~\ref{sec_model},
let $T$ denote the river network and assume that we have $n$
observations $\Xbf_1, \ldots,\Xbf_n \in\R^m$
at $m$ locations $\tbf= (t_1,\ldots, t_m) \in T^m$. Further, suppose
that the data are normalized to standard Pareto margins with cumulative
distribution function $1 - 1/x$ ($x\geq1$) and that
the vectors $\Xbf_{k}$ ($k=1,\ldots, n$) are independent copies
of a random vector $\Xbf$ in the max-domain of attraction of the
max-stable process $\eta_{\Gamma}(\tbf) =(\eta_\Gamma(t_1), \ldots
, \eta
_\Gamma(t_m))$.
This means that
%
\begin{equation}
\label{mda} \lim_{n\to\infty} n \P(\Xbf/ n \in A) =
V_{\Gamma,\tbf}(A),
\end{equation}
for any Borel subset $A \subset E$ which is bounded away from $\mathbf
0$ and which has zero $V_{\Gamma,\tbf}$ measure on its boundary; recall
the definition of the exponent measure in Section~\ref{sec_EVT}.

\subsubsection{Spectral estimation of \texorpdfstring{$\Gamma_\vartheta$}{Gamma_vartheta}}

The random vector $\eta_{\Gamma}(\tbf)$ follows a multivariate H\"
usler--Reiss distribution. Even though its multivariate densities are
not available, the densities of its exponent measure $V_{\Gamma,\tbf}$
have closed forms for any dimensions and we can apply the spectral
estimator proposed by \citet{eng2014}. Indeed, for large thresholds $u
> 0$ the convergence in \eqref{mda} justifies the approximation
%
\begin{equation}
\label{norm_cond} \P\bigl (\Xbf\in\D\mathbf x, \| \Xbf\|_1 > u \bigr) \approx-
\frac{\partial^m}{\partial x_1\cdots\partial x_m} V_{\Gamma,\tbf}(x_1,\ldots, x_m)\,\D
\mathbf x,
\end{equation}
where $\| \mathbf x\|_1 = \sum_{j=1}^m x_{j}$ $(\mathbf x\in E)$
denotes the $L_1$-norm,
and $V_{\Gamma,\tbf}(\{\mathbf x\in E: \|\mathbf x\|_1 > 1 \}) = m$.
Owing to the homogeneity of the exponent measure $V_{\Gamma,\tbf}$ in
Section~\ref{sec_EVT},
it suffices to specify the angular part of \eqref{norm_cond}, namely,
its \emph{spectral density} on the
positive $L_1$-sphere $S_{m-1} = \{\mathbf x\geq \mathbf
{0}:\|
\mathbf x\|_1 = 1\}\subset\R^m$ [\citet{col1991}]. \citet{eng2014} showed
that the spectral density of the H\"usler--Reiss exponent measure is
\begin{eqnarray}
g_\vartheta (\omega_1,\ldots,\omega_m ) =
\frac
{1}{\omega_1^2
\omega_2\cdots\omega_m (2\pi)^{(m-1)/2}
|\det\Sigma_\vartheta|^{1/2}} \exp \biggl( -\frac{1}2 \tilde{\bolds{
\omega}}^\T\Sigma_\vartheta^{-1}\tilde {\bolds {
\omega}} \biggr),\nonumber\\
 \eqntext{\bolds{\omega} \in S_{m-1},}
\end{eqnarray}
where $\tilde{\bolds{\omega}} = (\log(\omega_j / \omega_1) +
\Gamma_\vartheta(t_j,t_1) / 2: j=2,\ldots, m)^\T$ and $\Sigma
_\vartheta
\subset\R^{(m-1)\times(m-1)}$ is the covariance matrix
from Remark~\ref{rem_BR}(a) for $u = t_1$, that is,
%
\begin{equation}
\label{sigma} \Sigma_\vartheta= \tfrac{1}2 \bigl\{
\Gamma_\vartheta(t_i,t_1) + \Gamma
_\vartheta(t_j,t_1) - \Gamma_\vartheta(t_i,t_j)
\bigr\}_{2\leq
i,j\leq m}.
\end{equation}
Thus, denoting the index set of extremal observations by
$\mathcal I = \{k=1,\ldots, n: \| \Xbf_{k}\|_1 > u \}$, the spectral
estimator $\hat\vartheta_{\SPEC}$ of
$\vartheta$ is defined by
%
\begin{equation}
\label{spec_est} \hat\vartheta_{\SPEC} = \arg\max_{\vartheta\in\Theta}
\sum_{k\in
\mathcal I} \log g_\vartheta \bigl(
\Xbf_{k}/\| \Xbf_{k}\|_1 \bigr).
\end{equation}
The advantage of this estimator over composite likelihood counterparts
is that it uses a full likelihood and thus is fully efficient, thus
giving improved estimation of Brown--Resnick processes; see the
simulation study in \citet{eng2014}. Owing to the explicit form of the
spectral densities, this approach is feasible even for a large number
$m$ of locations.

\subsubsection{Censored estimation of \texorpdfstring{$\Gamma_\vartheta$}{Gamma_vartheta}}
\label{cen_est}

Conditioning on the norm of observations being large, as in \eqref
{norm_cond}, might introduce bias, since the limit distribution may
provide a poor density approximation to any of the $\Xbf_{k}$ that have
small individual components. To overcome this, \citet{wad2013} apply
\emph{censoring} to those components that do not exceed a fixed high
threshold. We adopt their approach, giving a new, simpler expression
for the censored likelihood, valid for any process with H\"usler--Reiss
margins, not just for stationary Brown--Resnick processes.

Similarly to the spectral estimation based on \eqref{norm_cond}, for
large thresholds $u > 0$ we have the approximation
%
\begin{eqnarray}
\label{max_cond} \P \Bigl(\Xbf\in\,\D\mathbf x, \max_{j=1,\ldots,m}
X_{j} > u \Bigr)\approx - \frac{\partial^m}{\partial x_1\cdots\partial x_m} V_{\Gamma,\tbf
}(x_1,
\ldots,x_m)\,\D\mathbf x.
\end{eqnarray}
Here, a multivariate observation is said to be extreme if at least one component
exceeds the threshold. For the likelihood contribution from an observation
$\Xbf= (X_1,\ldots,X_m)$ we distinguish two cases:
\begin{itemize}
\item
if at least one component exceeds the threshold, that is,
$X_{j}>u$ for all $j\in\mathcal K$ and $X_{j}\leq u$ for all $j\in
\mathcal K^C=\{1,\ldots, m\} \setminus\mathcal K$ for
a nonempty subset $\mathcal K\subset\{1,\ldots,m\}$, we compute the
likelihood $f_{\vartheta, \mathcal K}(\Xbf)$
by censoring all $\mathcal K^C$-components of the full likelihood
$f_{\vartheta, 1:m}(\Xbf)$.
We thus only use the information that those components are below the
threshold $u$, but not their exact values.
Without loss of generality, let $\mathcal K = \{1, \ldots,b\}$, for
some $b\in\{1,\ldots, m\}$. Then the censored likelihood is
%
\begin{eqnarray}
\label{eqn:likcontrib} f_{\vartheta, \mathcal K}(\mathbf x) &=& - \frac{\partial
^b}{\partial
x_1\cdots\partial x_b}V_{\Gamma,\tbf}(x_1,
\ldots, x_b, u,\ldots,u)
\nonumber
\\[-8pt]
\\[-8pt]
\nonumber
&=& \frac{1}{x_1^2 x_2\cdots x_b} \phi_{b-1} (\tilde{\mathbf {x}}_{2:b};
\Sigma_{2:b,2:b} ) \Phi_{m-b} ( \mu_C ;
\Sigma_C ),
\end{eqnarray}
where $\Sigma= \Sigma_\vartheta$ is the covariance matrix in \eqref
{sigma}, $\tilde{\mathbf{x}} = (\log{x_j} - \log{x_1} + \Gamma
_\vartheta
(t_j,t_1) / 2: j=1,\ldots, m)^\T\in\R^m$, and $\phi_{p}( \cdot,
\Psi)$ and $\Phi_{p}( \cdot, \Psi)$ denote the density and the
cumulative distribution function of a $p$-dimensional, zero-mean normal
distribution with covariance matrix $\Psi$.
We set $\phi_{0}$ to $1$ if $b=1$, and $\Phi_{0}$ to $1$ if $b=m$. The
conditional mean $\mu_C$ and
covariance matrix $\Sigma_C$ are
%
\begin{eqnarray}
\label{mu_C} \mu_C &=& \bigl(\log u - \log x_1 +
\Gamma_\vartheta (t_j,t_1) / 2
\bigr)_{j=b+1,\ldots, m} - \Sigma_{(b+1):m, 2:b} \Sigma_{2:b,2:b}^{-1}
\tilde{\mathbf {x}}_{2:b},
\\
\label{sigma_C}\Sigma_C & =& \Sigma_{(b+1):m,(b+1):m} - \Sigma
_{(b+1):m,2:b}\Sigma^{-1}_{2:b,2:b} \Sigma_{2:b,(b+1):m}.
\end{eqnarray}
In the case $b=1$, $\mu_C$ and $\Sigma_C$ are unconditional, that is,
the last summands
in the formulas above vanish.
The derivation of this new representation of $f_{\vartheta, \mathcal
K}$ can be found in \citet{Asadi.Davison.Engelke:2016:SM}.
\item
if none of the components exceeds $u$, that is, $\mathcal K =
\varnothing
$, then
the likelihood contribution is just the probability $ f_{\vartheta,
\mathcal K}(\mathbf x) = 1 - V_{\Gamma,\tbf}(\mathbf u)$ that $\Xbf$
lies entirely
below the threshold.
\end{itemize}

Let $\mathcal J = \{i=1,\ldots, n: \max_{k=1,\ldots,m} X_{i,k} > u \}
$ denote
the index set of observations extreme in the sense of \eqref{max_cond}
and, for each $i\in\mathcal J$,
let $\mathcal K_i$ be the index set of those components of $\Xbf_{i}$
that exceed $u$. Then, the censored estimator $\hat\vartheta_{\CENS}$
is obtained by maximizing the log-likelihood [\citet{thi2014}, Section~3]
%
\begin{equation}
\label{cens_est} \hat\vartheta_{\CENS} = \arg\max_{\vartheta\in\Theta}
\biggl[\bigl(n - |\mathcal J|\bigr) \log\bigl\{1 - V_{\Gamma,\tbf}(\mathbf u)\bigr\} +\sum
_{i\in\mathcal J} \log f_{\vartheta,\mathcal K_i} (\Xbf_{i} )
\biggr].
\end{equation}
This estimator has the advantage of using full likelihoods and reducing
potential bias by censoring components that might not yet have
converged, but the disadvantage of being slow when $m$ is large, since
the censored likelihood $f_{\vartheta,\mathcal K}$ then involves the
burdensome evaluation of high-dimensional normal distribution functions.

\subsubsection{Simulation study}
\label{sect:sims}

The two estimators $\hat\vartheta_{\SPEC}$ and $\hat\vartheta
_{\CENS
}$ use different data and will have different behavior for finite
sample sizes. We conducted a small simulation study to assess their
performance in a setting similar to our application. Details can be found
in \citet{Asadi.Davison.Engelke:2016:SM}. Both estimation procedures
work for the simulated data, even with a low number of observations;
only the extreme events contribute to the likelihoods. In simulated
data, the advantage of censoring cannot be seen, but it will reduce any
bias for real data. As also noted by \citet{eng2014} and \citet{ein2014},
the estimates of $\lambda_{\mathrm{Euc}}$ have larger variation than the
others. In fact, owing to a near-functional relationship between the
scale $\lambda_{\mathrm{Euc}}$ and the shape $\alpha$ of the fractal
variogram, these two parameters are strongly related in the range
considered here, and this near lack of identifiability gives highly
variable estimators of $\lambda_{\mathrm{Euc}}$.

\section{Extreme river discharges in the upper Danube basin}

\subsection{Data}
\label{data}
We used data for average daily discharges recorded at $m = 31$ German
gauging stations on $20$ rivers in the upper Danube basin, made
available by the Bavarian Environmental Agency (\url
{http://www.gkd.bayern.de}). The average discharges at these stations
range from around 20~m$^3$/s at high altitudes to around 1400m$^3$/s at
the most downstream station. The major part of the runoff in the basin
arises from the Alps, situated south of the Danube; see Figure~\ref
{Fig_Map}. The series at individual stations have lengths from 50 to
130 years, with 50 years of data for all stations from 1960--2009.
Originally, data were provided for 47 stations, but we excluded 16
stations which have very small discharges or whose largest discharges
are affected by hydroelectric installations or dampened by big lakes;
it might be possible to include these data by applying special
preprocessing techniques, but we have not explored this.

Exploratory analysis shows that around one-half of the annual maxima in
the basin occur in June, July and August. This agrees with the study of
floods in the Danube tributaries Lech and Isar by \citet{boe2006},
which shows that nearly all major floods in recent decades have
occurred in these three months; floods in this area are typically
caused by heavy summer rain. In order to eliminate temporal
nonstationarities and the effect of snow melt, we restrict our analysis
to these months. For $k= 1,\ldots, N$, we let $\Ybf_k = (Y_{1,k},
\ldots
,Y_{m,k})$ denote the daily mean discharge at the $m$ stations on day
$k$. The number of common measurements at all stations is thus $N = 50
\times92 = 4600$, that is, 50 years of 92 daily observations in the
summer months.

Seasonality and overall trend are the main sources of nonstationarity
in river flow data, but as we use only the summer month discharges, the
seasonality becomes negligible. National studies have concluded that
there are no significant trends in the extremes of stream flows in our
area of interest [\citet{kat2002, kun2005}], in agreement with our
exploratory analysis, so henceforth we treat our data as temporally stationary.

In addition to the time series of daily average discharges, we use a
digital elevation model to obtain the following geographical covariates
at each station: the latitude and longitude of both the station itself
and the weighted centroid of its sub-catchment, and catchment
attributes including its size, mean altitude and mean slope.

\subsection{Declustering}
\label{extraction}

Extreme discharges at a given station occur in clusters due to temporal
dependence, which must be removed for spatial modeling.
Moreover, a large value at an upstream station may cause a peak further
downstream a day or two later. These slightly shifted maximum values on
different rivers stem from the same event and should be treated as
dependent. In the framework of meteo\-rology, multivariate declustering
is used by \citet{taw1988}, \citet{col1991} and \citet{pal1999} to
extract independent ``storm events.'' We apply a similar technique to
obtain a set of independent flood events $\tilde\Xbf_{1}, \ldots,
\tilde\Xbf_{n} \in\R^m$ on the river network from the full time
series $\Ybf_{k}$ ($k= 1,\ldots, N$).

\begin{figure}

\includegraphics{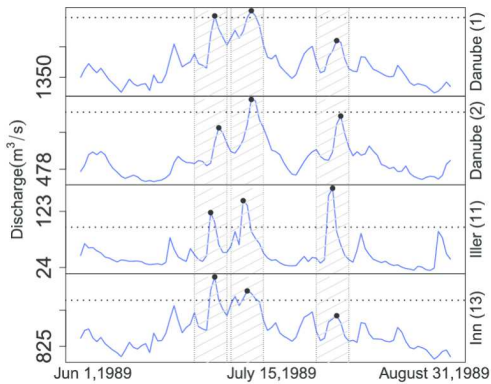}

\caption{Declustered flood events at four gauging stations. The grey
hatched areas are the
$p$-day time windows around flood events. Only events for which at
least one river exceeds its $90\%$
quantile (dotted horizontal lines) are shown. The black circles show
maxima for each river in each window.}
\label{Fig-DeclustPlot}
\end{figure}

In order to extract the flood events, we first identify nonoverlapping
windows of length $p$ days in each of the 50 summer periods. We replace
each observation by its rank within its series, and then consider the
day with the highest rank across all series, choosing this day randomly
if it is not unique. We then take a window of $p$ days centered upon
the chosen day, and form an event by taking the largest observation for
each series within this window. We delete the data in this window and
then repeat the process of forming events, stopping when no windows of
$p$ consecutive days remain. Figure~\ref{Fig-DeclustPlot} illustrates
this declustering procedure. In agreement with \citet
{kropp2010extremis}, our data suggest that flood events last no longer
than $9$ days, so we put $p = 9$; a sensitivity analysis showed that
our results are robust to this choice. For the $i$th time window, the
corresponding flood event $\tilde\Xbf_i$ is the $m$-dimensional vector
whose $j$th entry is the maximum discharge value at location $t_j$
within this window. This procedure yields a declustered time series of
$n = 428$ supposedly independent events $\tilde\Xbf_{i}$ from the $N =
4600$ summer measurements common to the 31 series.

\subsection{Marginal fitting}
\label{uni_est}

Before using the techniques from Section~\ref{sec_stat_mult}
to fit the multivariate dependence model, we
assess the univariate tail behavior at individual gauging stations,
obtaining the constants $a_{j,n},b_{j,n}$ and shape
parameters $\xi_j$ that allow us to normalize the margins to lie in the
standard Fr\'echet max-domain
of attraction, using \eqref{U_transform}.
The model $\tilde\eta_\Gamma$ in \eqref{eta_tilde} is a max-stable
stochastic
process on the whole river network $T$, so in order to make predictions
throughout $T$, we must allow the
norming constants and shape parameters to vary with covariates
that are easily obtainable even at locations without gauging stations
or find some other way to extend the model to the entire network, such
as kriging.

We fitted a generalized extreme value distribution \eqref
{exceed_pareto} to the tail of the declustered daily discharges at each
gauging station location $t_j$, estimating the extreme value parameters
$a_{j,n}$, $b_{j,n}$ and $\xi_j$. At each location we tested whether
the extremal behavior from any available earlier data changed relative
to the $50$ common years. In almost all cases there was no such change,
and we could use the longer series of independent events, declustered
using the procedure of Section~\ref{extraction}, for each station. For
the marginal fitting we use the independent events at gauging stations
and estimate the GEV parameters by maximizing the joint Poisson process
likelihood given in \eqref{PPP_density} in an independence likelihood
[\citet{Chandler.Bate:2007}].

We fitted and compared a variety of different models using this
technique, finally settling on a version of regional analysis, as
widely used in hydrological applications. The idea is similar to the
regionalization method of \citet{mer2005}, who predict high quantiles of
river flows using the catchment attributes of stations that are
``hydrologically'' close. Exploratory analysis suggests that for our
purposes the upper Danube basin can be split into four disjoint
regions: R1 contains eight stations in the southwest of the upper
Danube basin and has mid-altitude sub-catchments; R2 comprises five
stations in the Inn basin that are fed by precipitation in
high-altitude alpine regions; R3 contains 13 stations in the center of
the Danube basin that are fed by precipitation from regions with both
high and low altitudes; and R4 contains five stations with sources
north of the Danube. With $J_1,\ldots, J_4$ denoting the index sets of
stations in regions $R_1,\ldots, R_4$, we let for $j\in J_i$
$(i=1,\ldots
, 4)$
%
\begin{eqnarray}
\label{linear_model} \log(a_{j,n})& =&\sum_{k=1}^4
\alpha_{k}^{(i)} \log(P_{j,k}),
\nonumber
\\[-8pt]
\\[-8pt]
\nonumber
\log(b_{j,n})& =&\sum_{k=1}^4
\beta_{k}^{(i)} \log(P_{j,k}),\qquad \xi_j =
\xi^{(i)},
\end{eqnarray}
where $P_{j,1},\ldots,P_{j,4}$ are the latitude of the centroid,
the size, the mean altitude and the mean slope of the sub-catchment of
gauging station $j$. Likelihood ratio statistics were used to further
simplify the model, finally yielding a model with $28$ parameters,
compared to $93=3\times31$ parameters in the full model. Diagnostic
plots indicate a very satisfactory fit of the simpler model, which is
also strongly favored by the AIC. The estimated shape parameters and
their standard errors for the four regions are $0.030\ (0.025)$,
$0.145\ (0.034)$, $0.028\ (0.022)$ and
$0.294\ (0.045)$, suggesting that catchments influenced by mountain
regions tend to have heavier-tailed responses.

\begin{figure}

\includegraphics{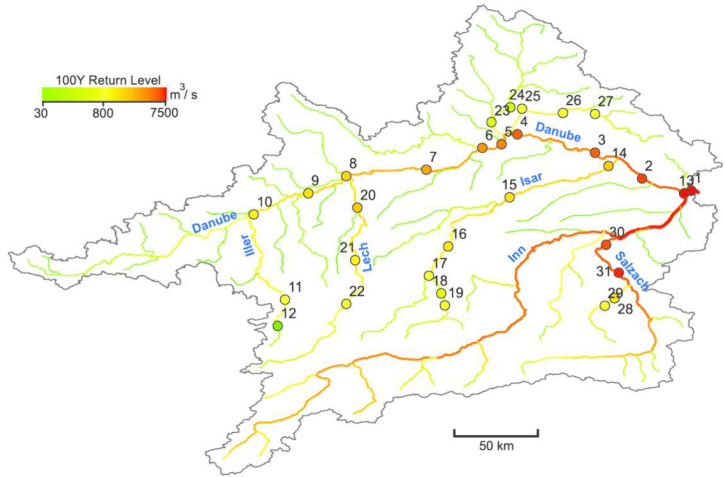}

\caption{100-year return levels for river flow (m$^3$/s), extrapolated
to the entire network $T$; the
colors of the points indicate the return levels at the $31$ numbered
gauging stations.}
\label{Fig_RetLevel}
\end{figure}

This model allows the extrapolation of the marginal fit to ungauged
locations on the network $T$, thereby enabling computation of return
levels throughout $T$; see Figure~\ref{Fig_RetLevel}. More details are
given in \citet{Asadi.Davison.Engelke:2016:SM}.

\subsection{Joint fitting}

\begin{figure}[t]

\includegraphics{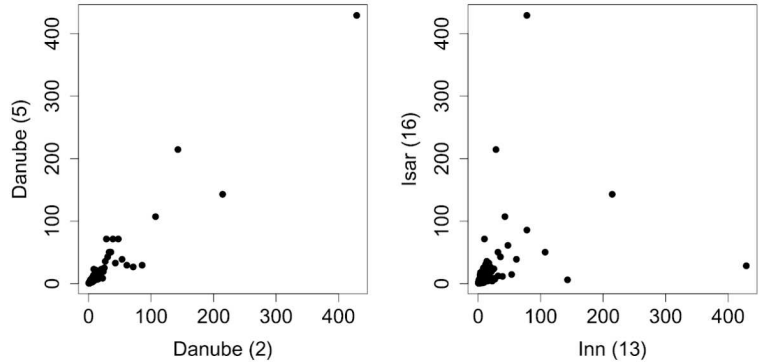}

\caption{Scatter plots of declustered discharges (normalized to the
unit Fr\'echet scale)
of two flow-connected stations (left)
and two flow-unconnected stations (right).}
\label{scatter}
\end{figure}

The generalized extreme value distributions constitute all possible
limits for univariate maxima, but the dependence structure of
multivariate extremes is infinite-dimensional, so we must first check
that the extreme discharges at different stations on the river network are
asymptotically dependent; if not, max-stable processes would not be
suitable models. \citet{kee2009a} note that the spatial dependence of
extreme river flows is much stronger than that of precipitation data,
since the former averages the latter and thus is less vulnerable to
small-scale variation, and standard diagnostics [\citet{col1999}] show
strong extremal dependence between all $31$ stations in our data.
Moreover, Figure~\ref{scatter} shows bivariate scatter plots of two
flow-connected and two flow-unconnected stations. In both cases, the
assumption of asymptotic dependence seems appropriate and, moreover, a
symmetric model for the tail dependence can be justified.

\begin{figure}[b]

\includegraphics{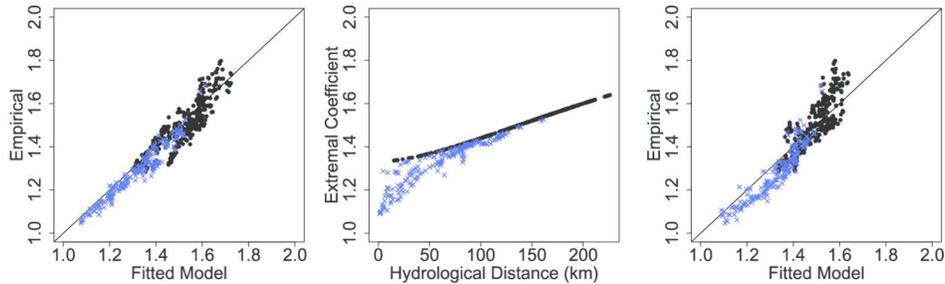}

\caption{Comparison of empirical estimates of extremal coefficients
found nonparametrically using
the madogram and those implied by different models, for all pairs of
gauging stations. Left:
madogram-based estimates and extremal coefficients $\hat\theta_{\HR}$
of the H\"usler--Reiss model,
estimated by fitting to independent events. Center: estimates using
$\hat\Gamma_3$ plotted against hydrological distance.
Right: madogram-based estimates and those from fitted joint model $\hat
\Gamma_3$. Those for
flow-connected pairs are blue crosses, and those for flow-unconnected
pairs are black circles.}
\label{EmpVSModel}
\end{figure}

The choice of a parametric subclass within the asymptotic dependence
models must be a good approximation to the infinite-dimensional
structure of multivariate max-stable distributions. Theorem~17 in \citet
{kab2009} gives some justification for the fitting of H\"usler--Reiss
distributions and Brown--Resnick type processes, which are essentially
the only possible limits of pointwise maxima of suitably rescaled and
normalized, independent, stationary Gaussian processes.

In order to assess whether the H\"usler--Reiss distribution
approximates the extremal dependence of our data well, we estimate the
extremal coefficient $\hat\theta$ as in \eqref{EC} for each pair of
locations using the madogram [\citet{coo2006}] based on summer
maxima. We then fit the bivariate H\"usler--Reiss distribution \eqref{biv_cdf}
to these data by a censored peaks-over-threshold approach and use
\eqref
{ECF} to compute a model-based extremal coefficient estimate $\hat
\theta_{\HR}$. The left panel of Figure~\ref{EmpVSModel} suggests that
the H\"usler--Reiss model provides an excellent overall approximation
to the bivariate extremal dependence structure of the discharge data,
albeit with slight overestimation of dependence at longer distances for
flow-unconnected pairs.

We compare four overall models for the dependence kernel $\Gamma$:
\begin{itemize}
\item the stationary variogram based on Euclidean distances with
anisotropy matrix R as in \eqref{aniso_mat},
\[
\Gamma_1(s,t) = \lambda\bigl\| R\cdot(s - t) \bigr\|^\alpha,\qquad \lambda>
0, \alpha\in(0,2], \beta\in[\pi/4,3\pi/4], c > 0;
\]
\item a variogram using the transformation $H$ to hydrological locations,
\begin{eqnarray}
\Gamma_2(s,t) = \lambda\bigl\| R\cdot\bigl\{H(s) - H(t)\bigr\}
\bigr\|^\alpha,\nonumber\\
 \eqntext{\lambda> 0, \alpha\in(0,2], \beta\in[\pi/4,3\pi/4], c > 0;}
\end{eqnarray}
\item a variogram that includes the hydrological properties of the
river network for flow-connected locations, corresponding to \eqref
{dep_kernel},
\[
\Gamma_3(s,t) = \lambda_{\mathrm{Riv}} \Gamma_{\mathrm{Riv}}(s,t) +
\lambda _{\mathrm{Euc}} \bigl\| R\cdot\bigl\{H(t) - H(s)\bigr\}\bigr \|^\alpha,
\]
whose six parameters are given in \eqref{six_par}; finally,
\item
we also consider the previous model without anisotropy,
\begin{eqnarray}
\Gamma_4(s,t) = \lambda_{\mathrm{Riv}} \Gamma_{\mathrm{Riv}}(s,t) +
\lambda _{\mathrm{Euc}} \bigl\| H(t) - H(s)\bigr \|^\alpha,\nonumber\\
 \eqntext{\lambda_{\mathrm{Riv}},
\lambda_{\mathrm{Euc}} > 0, \tau> 0, \alpha\in(0,2].}
\end{eqnarray}
\end{itemize}
The weights in $\Gamma_{\mathrm{Riv}}$ are computed according to \eqref
{weights_def} using a digital elevation model.

In Section~\ref{extraction} we extracted $n = 428$ independent
multivariate flood events $\tilde\Xbf_1,\ldots,\tilde\Xbf_n$, whose
univariate extremal behavior was analyzed in Section~\ref{uni_est}. In
order to fit the multivariate dependence structure, we use the marginal
empirical distribution functions to transform the distribution at each
gauging station to standard Pareto, and denote the resulting data by
$\Xbf_1,\ldots, \Xbf_n$. We fit the functions $\Gamma_1,\ldots,
\Gamma
_4$ for the negative definite kernel
in $\eta_\Gamma$ to these data using the inference procedures described
in Section~\ref{sec_stat_mult}, first obtaining
the spectral estimate $\hat\vartheta_{\SPEC}$ in \eqref{spec_est} by
grid search on the parameter space $\Theta$, and then using this as an
initial value for the more demanding computation of the censored
estimate $\hat\vartheta_{\CENS}$ in \eqref{cens_est}. It would be
preferable to fit the univariate margins and the dependence structure
simultaneously, but here this is infeasible since the optimization for
the dependence structure is very time intensive.

The maximized log-likelihoods corresponding to $\Gamma_1,\ldots,
\Gamma
_4$ are $-6629.17$, $-6161.86$, $-5907.49$ and $-5915.97$; $\Gamma_3$
has six parameters, and the others all have four parameters. The use of
hydrological distances for $\Gamma_2,\Gamma_3, \Gamma_4$ gives a huge
improvement over the use of Euclidean distances in $\Gamma_1$, and
adding the component $\Gamma_{\mathrm{Riv}}$ for flow-connected dependence
means that $\Gamma_3$ is much better than $\Gamma_2$. The drop from
$\Gamma_3$ to $\Gamma_4$ shows that the anisotropy matrix $R$ also
contributes to the good fit of the model based on $\Gamma_3$.

The center and right panels of Figure~\ref{EmpVSModel} (recall also the
right panel of Figure~\ref{ECF_Euc}) compare the extremal coefficients
obtained with the madogram and those implied by the fitted model
$\Gamma
_3$. The center panel shows that the latter do not lie on a smooth
curve; flow-connected pairs at the same distance can have different
extremal coefficients, depending on where the two stations lie on the
network, because the river dependence kernel $\Gamma_{\mathrm{Riv}}$ is
nonstationary, unlike those based on simple meteorology. Overall there
is a fairly good fit, though the model tends to slightly understate
dependence at short hydrological distances and to overstate it at long ones.

The parameter estimates $\hat\vartheta_{\CENS}$ are $\hat\lambda
_{\mathrm{Riv}}=0.73\ (0.07)$, $\hat\lambda_{\mathrm{Euc}} =1.93\times
10^{-4}  (0.75\times10^{-4})$, $\hat\tau=839\ (280)$ km, $\hat
\alpha
=1.75\ (0.08)$, $\hat\beta=1.10\ (0.11)$ and $\hat c=0.64\ (0.08)$, with
standard errors in parentheses obtained from 100 nonparametric bootstrap
simulations. The high uncertainty for $\hat
\lambda_{\mathrm{Euc}}$ was mentioned when discussing the simulation study;
it does not translate into high variation of the fitted model.

The fitted weights $\hat\lambda_{\mathrm{Riv}}$ and $\hat\lambda_{\mathrm{Euc}}$ cannot be compared directly, because the variogram $\Gamma_{\mathrm{Euc}}$ is unbounded and thus does not have
a natural normalization. The influences of the river and the Euclidean
dependence kernel on the overall extremal dependence between two
flow-connected points $s,t\in T$ can be measured by $\hat\Gamma_{\mathrm{Riv}}(s,t) / \hat\Gamma_3(s,t)$ and $\hat\Gamma_{\mathrm{Euc}}(s,t) /
\hat
\Gamma_3(s,t)$, respectively. In fact,
for certain pairs of stations the river dependence kernel is dominant,
whereas for others the Euclidean kernel has a stronger
influence on the extremal dependence.
The parameter $\hat\tau$ is the scale for dependence along the river;
as expected, this dependence is very strong, decreasing to zero only after $\hat
\tau= 839$ km. The shape parameter $\hat\alpha$ describes how local
the influence of spatial meteorological events on river flows is; note
that $\hat\alpha=1.75$ is much larger than in applications on extreme
precipitation, confirming the observation of \citet{kee2009a} that
extreme river flows exhibit stronger spatial dependence due to an
averaging effect. The parameters $\hat\beta$ and $\hat c$ describe the
anisotropy of meteorological dependence, since the transformation
$R(\hat\beta,\hat c)$ dilates the space in direction $(\sin\hat
\beta
, \cos\hat\beta)$ by $\hat c$. As $\hat c < 1$, extremal dependence
is increased in this direction, which corresponds approximately to the
planar vector $(2,1)$. Thus, in terms of hydrological distance, two
stations that are 64 km apart in a direction roughly parallel to the
Alps have the same dependence as two stations that are 100 km apart
perpendicular to the Alps. In view of the orientations of the
catchments and the blocking effect that the Alps have on weather
systems, this seems quite plausible.

\subsection{Higher-order properties}

\begin{figure}

\includegraphics{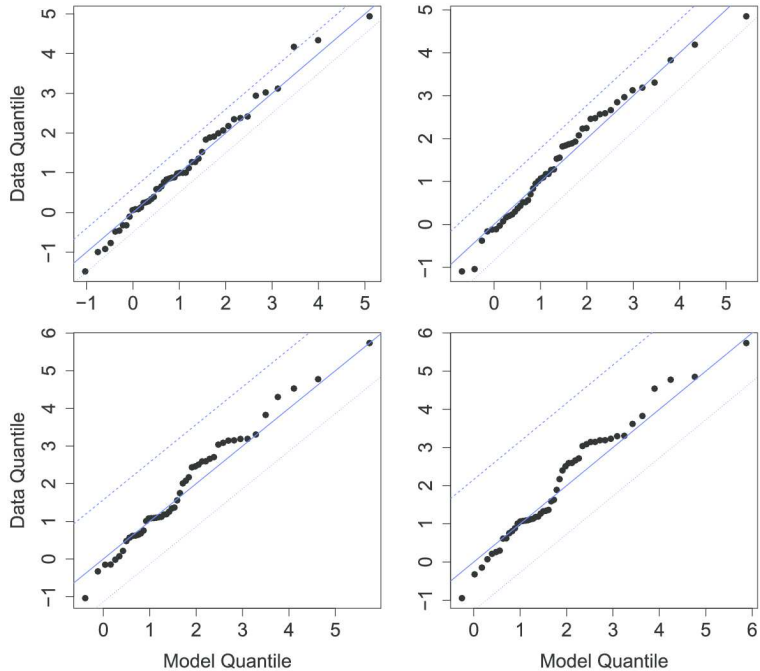}

\caption{QQ-plots (Gumbel scale) of observed groupwise yearly maxima and
theoretical values from the fitted model, for groups of $3$ (top left), $5$
(top right), $15$ (bottom left) and all $31$ (bottom right) stations.
Dashed lines and dotted lines correspond to values for complete
independence and
complete dependence, respectively, and the solid line corresponds to
the fitted model.}
\label{fig:QQ_groups}
\end{figure}

Figure~\ref{EmpVSModel} shows how the max-stable model $\eta_{\Gamma
_3}$ fits the bivariate extremal features of the data. In practice,
higher-order properties such as multivariate
exceedance probabilities are also of interest, and to check these we
randomly choose groups of $3$, $10$, $15$ and $31$ stations and compute
the quantiles of their observed group maxima, suitably rescaled
[cf. \citet{dav2012}]. Figure~\ref{fig:QQ_groups}, which compares these
quantiles with the theoretical values derived from the fitted model,
shows that the model captures even high order structures of the data
very well. Moreover, the comparison of observed quantiles to those
corresponding to complete independence and complete dependence
underlines the importance of proper dependence modeling.

A joint extremal model allows the estimation of the risk of
simultaneous exceedances of high thresholds at multiple
locations. More precisely, we can use equation \eqref{mult_exceed} to
approximate these probabilities as a function of the univariate extreme
value parameters
and the exponent measure $V$ of the dependence model. For three
stations $\mathbf t = (t_1,t_2,t_3) \in T^3$, the exponent measure
for our model is $V_{\Gamma, \tbf}$ as in \eqref{HR}. Let $q_{j,p}$ be
the $p$-quantile of the distribution of daily discharges at station~$t_j$. The probability of a flood that exceeds the respective
$p$-quantiles at all three stations in the same summer can be
approximated by
%
\begin{eqnarray}
\label{exceed_prob} &&K \P \bigl( X(t_j) > q_{j,p}; j = 1,2,3
\bigr)
\nonumber
\\[-8pt]
\\[-8pt]
\nonumber
&&\qquad\approx V_{\hat\Gamma_3, \tbf} \Biggl\{ \prod_{j=1}^3
\biggl( \biggl(1 + \hat\xi_j \frac{q_{j,p} -
\hat
b_{j,n}}{\hat a_{j,n}}
\biggr)_+^{1/\hat\xi_j}, \infty \biggr) \Biggr\},
\end{eqnarray}
where $K$ is the mean number of multivariate events per year. The
estimates for the shape and scale parameters are taken from the fitted
covariate model in \eqref{linear_model}, so this multivariate
exceedance probability, and others for more complex events, can be
computed for any locations, even ungauged, on the river network. To
compare the model with empirical data, we randomly choose $500$ out of
the $31 \choose3$ possible triplets of gauging stations and evaluate
\eqref{exceed_prob} for different values of $p$ close to $1$. The mean
relative absolute differences of these model probabilities and their
empirical counterparts are $15\%$ for $p = 0.95$, $14\%$ for $p =
0.97$, $19\%$ for $p = 0.99$, and $31\%$ for $p = 0.995$; the empirical
counterparts are highly variable, since they are based on very few events.

\section{Discussion}

The approach described above was used to fit other max-stable
processes, such as the extremal-$t$ or Schlather models, but we found
that the Brown--Resnick model was the best of those fitted; perhaps
this is not surprising, since this model is flexible and allows
independent extremes at long distances, unlike the Schlather model, for example.

\citet{kee2009a,kee2009,kee2013} describe an alternative approach to
modeling joint flooding that allows the possibility of asymptotically
independent extremes through the fitting of the \citet{hef2004} model.
This can handle large-scale problems, but has the drawback of not
treating the variables symmetrically, and it is not clear whether it
corresponds to a well-defined joint model. In those papers, it is
important to allow for asymptotic independence because the data arise
from rivers that may be quite unrelated, whereas stronger dependence
might be anticipated in a single river network, as in the present
paper. Moreover, our approach uses the known structure of the river
networks, which should provide better dependence modeling.

Finally, the ideas suggested here might be extended to similar problems
for which Euclidean geometry does not seem natural, such as the
transmission of earthquake shocks along fault lines, or communication
networks, though it would then be important to allow for flows in
different directions. In some applications it might be useful to
include the relative timings of extremes at different nodes of the network.

\section*{Acknowledgments}
We thank Jonathan Tawn, Hansjoerg Albrecher, Marianne Milano and the
editorial team for helpful remarks.

\begin{supplement}[id=suppA]
\stitle{Supplement to ``Extremes on river networks''\\}
\slink[doi,text={10.1214/\break 15-AOAS863SUPP}]{10.1214/15-AOAS863SUPP} 
\sdatatype{.pdf}
\sfilename{aoas863\_supp.pdf}
\sdescription{The
supplementary material contains the following: a~PDF document
containing the derivation of the new likelihood representation
mentioned in Section~\ref{cen_est}, results of the simulation study mentioned
in Section~\ref{sect:sims}, and additional details germane to
Section~\ref{uni_est}; and R
code and data files to reproduce the data analysis and figures.}
\end{supplement}

%



\printaddresses
\end{document}